\documentclass[%
 reprint,
 amsmath,amssymb,
 aps,
]{revtex4-2}

\usepackage{graphicx}
\usepackage{dcolumn}
\usepackage{bm}
\usepackage[dvipsnames]{xcolor}
\usepackage{pifont}
\usepackage{float}

\begin{document}

\preprint{APS/123-QED}

\title{Packing fraction related transport in disordered quantum dot arrays}


\author{K. Eshraghi\thanks{Program in Materials Science, University of California, San Diego, La Jolla, CA, 92093}$^{1}$}

\author{S. Natani\thanks{Department of Mechanical Engineering, University of California, San Diego, La Jolla, CA, 92093}$^{2}$}


\author{P. Bandaru\thanks{Program in Materials Science, University of California, San Diego, La Jolla, CA, 92093}$^{1}$\thanks{Department of Mechanical Engineering, University of California, San Diego, La Jolla, CA, 92093}$^{,2}$}


\affiliation{Program in Materials Science, University of California, San Diego, La Jolla, CA, 92093
}%
\affiliation{Department of Mechanical Engineering, University of California, San Diego, La Jolla, CA, 92093
}%

\date{\today}

\begin{abstract}
Models to describe electrical conduction in quantum dot (QD) constituted films often overlook the effects of geometric disorder. We address related issues by examining the influence of the QD packing fraction (PF) on the charge transport and transmission in QD arrays. Using transfer matrix based algorithms and Monte Carlo simulations, we quantify the transmission across disordered QD assemblies. Our results indicate a critical packing fraction ($PF_c $) of $\sim$ 0.64, marking a transition from a non-conducting to a conducting state, aligning well with experimental observations and analytical predictions. This study enhances the understanding of transport in QD arrays, with implications for designing efficient electronic devices based on disordered nanoscale systems.

\end{abstract}

\maketitle


While the size-dependent tunability\textsuperscript{1,2,3} of the energies in quantum dots (QD) offers an attractive degree of freedom for electronic device design, the consequent electrical conduction within larger scale QD-constituted assemblies and arrays has not yet been well understood.\textsuperscript{4} For instance, charge transport in QD films has been characterized\textsuperscript{5,6,7,8,9 }through nearest-neighbor-hopping\textsuperscript{7,8 }(NNH) as well as variable-range-hopping (VRH)\textsuperscript{6,9 }based models. While such models have been found to be useful in predicting the temperature\textsuperscript{6,7,9} and  bias voltage\textsuperscript{6,7,9} dependence of electrical current in QD assemblies, the energy scales related to the necessary electronic coupling\textsuperscript{ 7, 10} are yet based on the \textit{individual} characteristics of the constituent QDs, \textit{e.g.} with an assumed QD diameter: \textit{2R,} and an average fixed neighbor spacing: \textit{s}, \textit{etc.,} The treatment of size disorder in the hopping conduction models is not typically accurate\textsuperscript{ 7} to beyond an associated value of the quantized energy (\textit{E\textsubscript{z}) }obtained from the individual QDs, \textit{i.e. }with $\delta E_z/E_z \sim 2\delta R/R$.

Moreover, practically, when QDs are spin-coated\textsuperscript{10} or dispersed\textsuperscript{11} onto a substrate into a film for device application, it is very likely that geometrical disorder, \textit{e.g.} sub-optimal array packing and the related presence of vacancy defects, \textit{etc.} would be present. Such disorder in the film would lead to variation in  the width of energy barriers between the QDs as well as misalignment of the discrete energy levels related to the individual QDs. Here, we aim to broaden the understanding of the influence of disorder through quantifying its impact on electrical conduction in a QD film, with broad implications to the modeling of transport in higher dimensional systems constituted from lower dimensional moieties. 

One way to consider the contribution of spatially separated QDs placed in an array, to passing an externally measured current, is through varying the packing fraction (\textit{PF}) of QDs - which could be a measure of the overall disorder. As the \textit{PF }is increased(/decreased), the extent to which the QDs couple to one another would be enhanced (/diminished). It may then be expected that at a critical threshold\textsuperscript{12,13} value, say $PF_c$, there would be a transition from an electrically disconnected to a connected state in the QDs constituting the film, subsequent to which a measurable current through the assembly would be obtained. Such phenomenology should be manifested over a range of scales, from QD incorporated one-dimensional arrays \textit{to} thin films and extending to three-dimensional assemblies.

We initially consider charge transport in the QD constituted array: Fig. 1(a) as mediated through discrete energy levels: \textit{E\textsubscript{z }, }between two adjacent QDs \textit{via} resonant tunneling based transmission, and modeled through a tunneling probability: \textit{T(E\textsubscript{z})}. The QDs were placed on a two-dimensional hexagonal lattice, of \textit{net} area: $A_{Lattice}(=R*(2*N_x-2)\sqrt3R*(N_y-1) )$ with \textit{N}\textsubscript{x} and \textit{N}\textsubscript{y} as the number of sites in the orthogonal \textit{x} and \textit{y} directions, respectively, with each lattice site considered as a circle of radius \textit{R} (say, $\sim$ 2.5 nm) corresponding to the plane-projected area of a single spherical QD. As the number of vacant lattice sites/vacancies increases, the \textit{PF} $(=A_{QD}/A_{Lattice})$ decreases, where \textit{A\textsubscript{QD} } $(=Q*\pi R^2)$ is the sum of the projected areas of the QDs, with \textit{Q} being the number of QDs. 

\begin{figure}[H]
    \centering
    \includegraphics[width=1\linewidth]{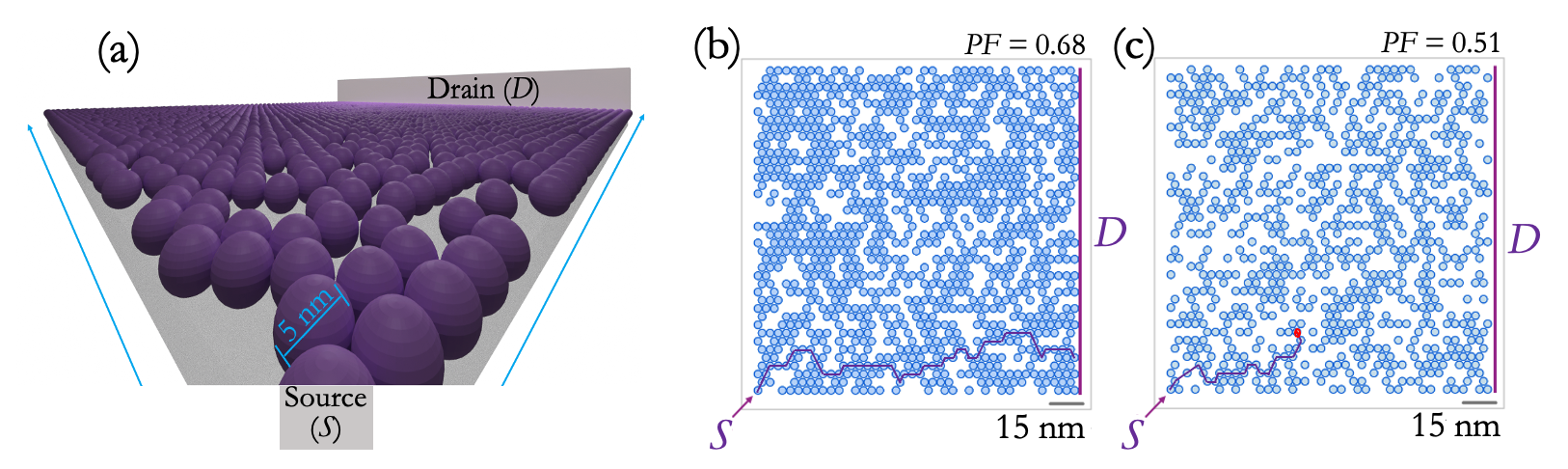}
    \caption{Modeling transmission through an array of QDs with varied amounts of disorder. (a) The transmission probability, $T(E_z)$, was monitored along possible charge transfer paths between a source (\textit{S}) QD and the drain (\textit{D}). (b) and (c) indicate sample configurations of the QD layer at a \textit{PF} of (b) 0.68 and (c) 0.51. The former (/latter) case leads to a successful (/unsuccessful) path between the \textit{S}  and the \textit{D}. }
    \label{1}
\end{figure}

The vacancies within the QD array were modeled to function as potential energy barriers of height \textit{E\textsubscript{b} (e.g. }a typical value being $\sim$ 5.07 eV, for, say, 5 nm diameter PbSe QDs\textsuperscript{11}) to transport, hindering charge carrier motion. The \textit{T(E\textsubscript{z})} was then obtained through a transfer matrix algorithm (\textit{TMA})\textsuperscript{ 14, 15, 16} applied to an electrical potential landscape (\textit{EPL}), comprised of a sequence of quantum wells and barriers encountered by a charge carrier within the QD array/assembly. 

An estimate of \textit{T(E\textsubscript{z}) }for any given \textit{EPL} was made by first considering a probability matrix, $M(\varepsilon,E_z)$: Eq. (1), defined as a function of both the applied electric field, $\varepsilon$, causing the charge motion, as well as the \textit{E\textsubscript{z }}. Each \textit{EPL} was segmented into \textit{n }distinct one-dimensional regions, with \textit{n} large enough to ensure independence of the estimates. 

\begin{equation}
M(\varepsilon, E_z) = \prod_{i=1}^{n} M_{i_c}(\varepsilon, E_z) \, M_{i_s}(\varepsilon, E_z)
\end{equation}

The probability of charge motion through each \textit{EPL} was estimated through compounding of the individual tunneling probabilities through the (i)  $M_{i_c}$ – along discrete steps of width $\Delta x$ in the \textit{i\textsuperscript{th} }potential barrier, for a wavevector (\textit{k\textsubscript{i}) } as related through Eq. (2)\textbf{, }\textit{i.e. }

\begin{equation}
M_{i_c}(\varepsilon,E_z)  = \begin{pmatrix}
e^{i k_i(\varepsilon, E_z) \Delta x} & 0 \\
0 & e^{-i k_i(\varepsilon, E_z) \Delta x}
\end{pmatrix}
\tag{2}
\end{equation}

as well as the (ii)  $M{_{i_s}}$ - across the interface of the \textit{i\textsuperscript{th} }and \textit{(i+1)\textsuperscript{th}} potential barriers\textsuperscript{ 14,15} as in Eq. (3)\textbf{: } 

\begin{equation}
M_{i_s}(\varepsilon,E_z) = \frac{1}{2} \begin{pmatrix}
1 + \frac{k_i(\varepsilon,E_z)}{k_{i+1}(\varepsilon,E_z)} & 1 - \frac{k_i(\varepsilon,E_z)}{k_{i+1}(\varepsilon,E_z)} \\
1 - \frac{k_i(\varepsilon,E_z)}{k_{i+1}(\varepsilon,E_z)} & 1 + \frac{k_i(\varepsilon,E_z)}{k_{i+1}(\varepsilon,E_z)}
\end{pmatrix}
\tag{3}
\end{equation}

Both the reflection coefficient (\textit{r}) as well as the transmission coefficient (\textit{t}), related to the charge transport were then obtained from the off-diagonal $M_{12}(=|r^2|/|t^2|)$ components of $M(\varepsilon,E_z)$.

From\textsuperscript{15 } $|t^2|+|r^2|=1$ we obtain the tunneling probability, $T(=|t^2|)(\equiv T(E_z))$ for a charge propagating through the \textit{EPL}. It is expected that as the number of vacancies (/\textit{PF}) increases (/decreases) that the \textit{T(E\textsubscript{z}) }would decrease. Our approach, where a \textit{TMA} is utilized to quantify the \textit{T(E\textsubscript{z}) }through a one-dimensional potential landscape\textsuperscript{14 }- associated with an arrangement of QDs\textsuperscript{17}, explicitly considers the prevalence of vacancies in a QD film and is capable of being extended to higher-dimensional arrangements of QDs, as will be indicated later.

A related Monte-Carlo (MC) simulation was deployed to probe geometrical disorder within the array, with the aim of significantly improving on previous\textsuperscript{18} work which was limited in scope, \textit{e.g.} to investigating \textit{positional }disorder – where the QDs were displaced up to 0.3 nm following a specific statistical distribution\textsuperscript{19}, or through a small change in the packing fraction\textsuperscript{20}. When size disorder was investigated\textsuperscript{21} in randomly close-packed QD assemblies, with QD diameters varied in the range of 3 nm - 8 nm, implying a spread in the lowest unoccupied molecular orbital (LUMO) energy positions of the individual QDs, a threshold to carrier transport was observed when the energy spread was below a certain value. However, the inevitable presence of vacancy defects with respect to the packing fraction of the assembly was not considered, as emphasized in our study.

We address aspects related to the shortcomings of the earlier work/s through investigating in detail the influence of QD vacancies as well as the related \textit{PFs} over a much larger range. Here, the impact of geometrical disorder on the charge transport was considered through hundreds of randomly generated\textsuperscript{22} instances of a two-dimensional 40 x 40 QD array. Such a representative array size was chosen to facilitate plausible comparison with experimental results\textsuperscript{13}, as will be indicated later. We have also ensured that the presented results were independent of the array size through investigating a wide range of array sizes. In each tested 40 x 40 monolayer, the \textit{number} as well as the \textit{position} of vacant lattice sites was randomly varied over a range of \textit{PFs }from 0.01 to 0.91, with the minimum (/maximum) value corresponding to all lattice sites vacant (/optimal packing\textsuperscript{23}). To model the motion of a charge within any given disordered monolayer, we incorporate the aspect that from any particular lattice site, numerous paths of charge transport would be available, representing lateral fluctuations\textsuperscript{24} of the conduction paths. For instance, Figs. 1(b), and 1(c), indicate sample tested trajectories through specific instances/\textit{PF}s of a disordered QD monolayer.

\begin{figure}[H]
    \centering
    \includegraphics[width=1\linewidth]{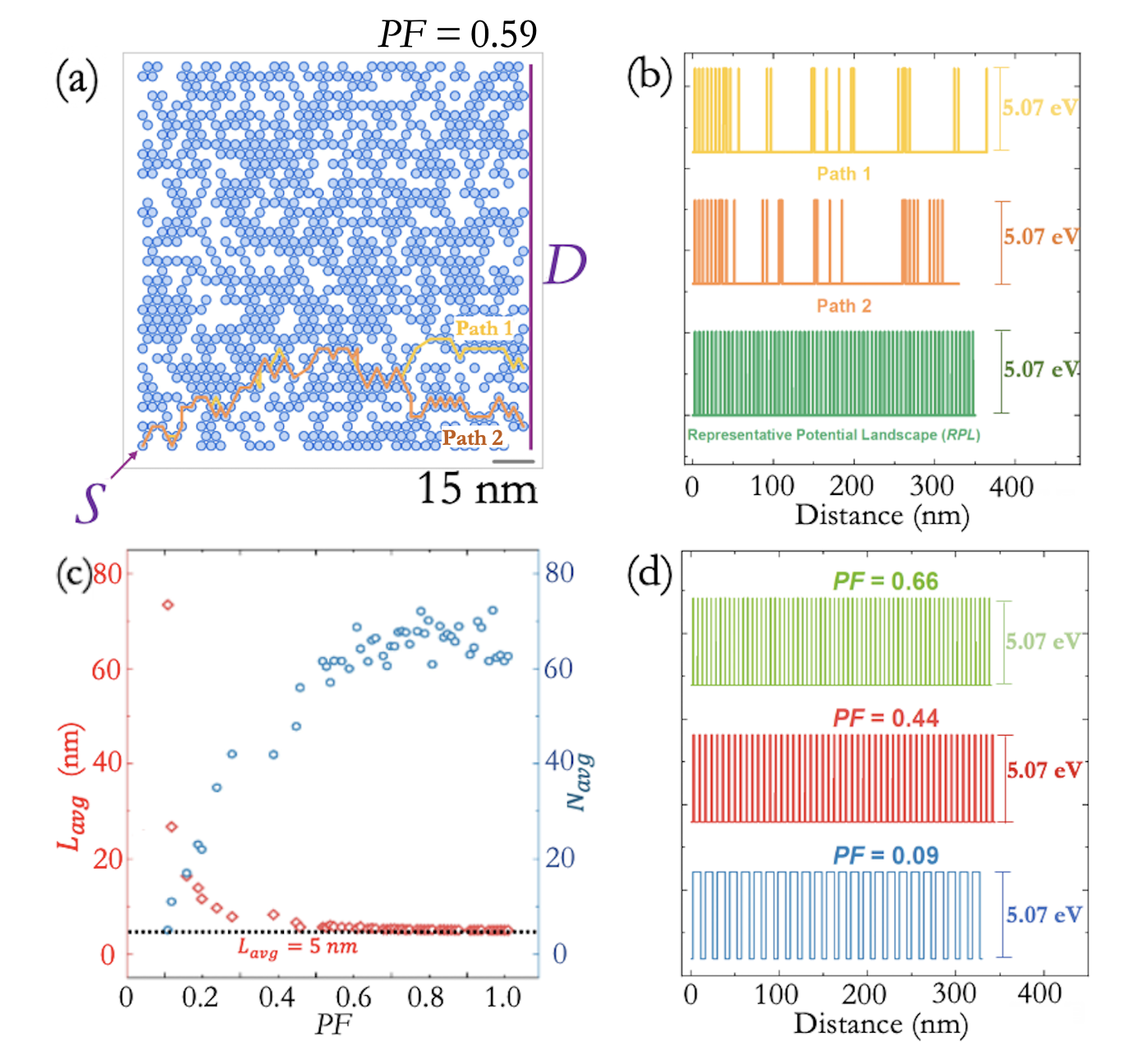}
    \caption{(a) Two representative successful sample paths through a randomly configured 40 x 40 QD monolayer with a \textit{PF} of 0.59. (b) The two successful paths in (a) were interpreted in terms of electrical potential landscapes (\textit{EPL},) and combined into a uniformly periodic representative potential landscape \textcolor{ForestGreen}{(\textit{RPL}, green)} consisting of $N_{avg}$ number of barriers, each of width $L_{avg}$ at a \textit{PF} of 0.59. (c) The mean jump length \textcolor{Red}{($L_{avg}$ , left)} and the mean number of jumps \textcolor{Periwinkle}{($N_{avg}$ , right)} between QDs in an array as a function of the \textit{PF}. (d) \textit{RPLs} for selected \textit{PF}s - an increasing \textit{PF} implies a decreasing barrier width for any given \textit{RPL}. }
    \label{2}
\end{figure}

We simulated millions of such possible trajectories, leading towards the approximation of a plausible path that may be selected by the charge – see Fig. 2(a). We first define \textit{$N_S$} as the number of successful paths in the entire set of simulated paths, related to the charge proceeding all the way from a single source QD to the drain – as determined through the following three criteria, \textit{i.e., }the path (i) may not intersect the same QD more than once, (ii) is forwardly biased (due to a potential difference, say, of $\sim$ 0.2 V between the \textit{S }and the \textit{D }– a typical value used in past experiments\textsuperscript{6} probing charge transport in QD assemblies), and (iii) involves tunneling \textit{only }between nearest neighbor QDs. Any path which did not satisfy all of the above criteria was considered \textit{unsuccessful} and does not contribute to the \textit{T(E\textsubscript{z}) }. After discarding such unsuccessful paths, the number of jumps (\textit{N\textsubscript{i}}), the total pathlength, \textit{ P\textsubscript{i,}} (determined by summing the net distance covered in each individual jump) for the \textit{i\textsuperscript{th}} successful path\textit{, }as well as the mean jump length $L{_{avg_i}}=P_i/N_i$, were computed for each successful path. Fig. 2(a) shows examples of two successful paths through a particular array configuration at a specific \textit{PF}. Each of these paths corresponds to a unique modulation of the electrical potential related to the charge motion: Fig. 2(b), as manifested through the related \textit{EPL}.

Furthermore, a \textit{mean }jump length $(L_{avg})(=\frac{\sum_{i=1}^{N_S}L{_{avg_i}}}{N_S})$ and mean number of jumps $(N_{avg})(=\frac{\sum_{i=1}^{N_S}N{_{avg_i}}}{N_S})$ are defined. It was observed that the \textit{L\textsubscript{avg} }was found to decrease with increasing \textit{PF: }Fig. 2(c)\textbf{,} eventually reaching an asymptote at $\sim$ 5 nm (equal to the center to center spacing of the QDs in the assembly.) In contrast, the \textit{N\textsubscript{avg} }increases with the \textit{PF}, implying an increase in the number of QDs involved in the transport. Subsequently, a uniformly periodic representative potential landscape (\textit{RPL}), consisting of \textit{N\textsubscript{avg }}number of barriers, each of width \textit{L\textsubscript{avg} }was generated for any given \textit{PF: }Fig. 2(d). The \textit{TMA} was then applied to the deduced \textit{RPL}, resulting in a \textit{PF} dependent transmission probability \textit{T(E\textsubscript{z}, PF) } through the \textit{entire }40 x 40 unit cell constituted array. At each \textit{PF, }the obtained \textit{T(E\textsubscript{z}) }results were averaged: $\overline{T}=\frac{\int_{0}^{E_b}T(E_z)dE_z}{E_b}$ over the entire range of\textit{ E\textsubscript{z} }present in the QW, yielding a defined $\overline{T}$, as indicated on \textit{the left axis} of Fig. 3. A threshold value\textit{, i.e., }a \textit{PF\textsubscript{c} }of $\sim$ 0.64 is indicated, where below (/above) the \textit{PF\textsubscript{c}}, the relatively large (/small) number of vacancies in the QD monolayer imply that the path a charge would traverse the layer would incorporate wider (/narrower) potential barriers - as indicated by the lower (/upper) \textit{RPLs }in Fig. 2(d) - leading to a diminished (/enhanced)  $\overline{T}$.

Such a \textit{PF\textsubscript{c }}value was found to be in close agreement with analytically determined\textsuperscript{25 }values of $\sim$ 0.65 of the critical percolation thresholds for a two-dimensional hexagonal lattice, as shown in Fig. 3.

\begin{figure}
    \centering
    \includegraphics[width=1\linewidth]{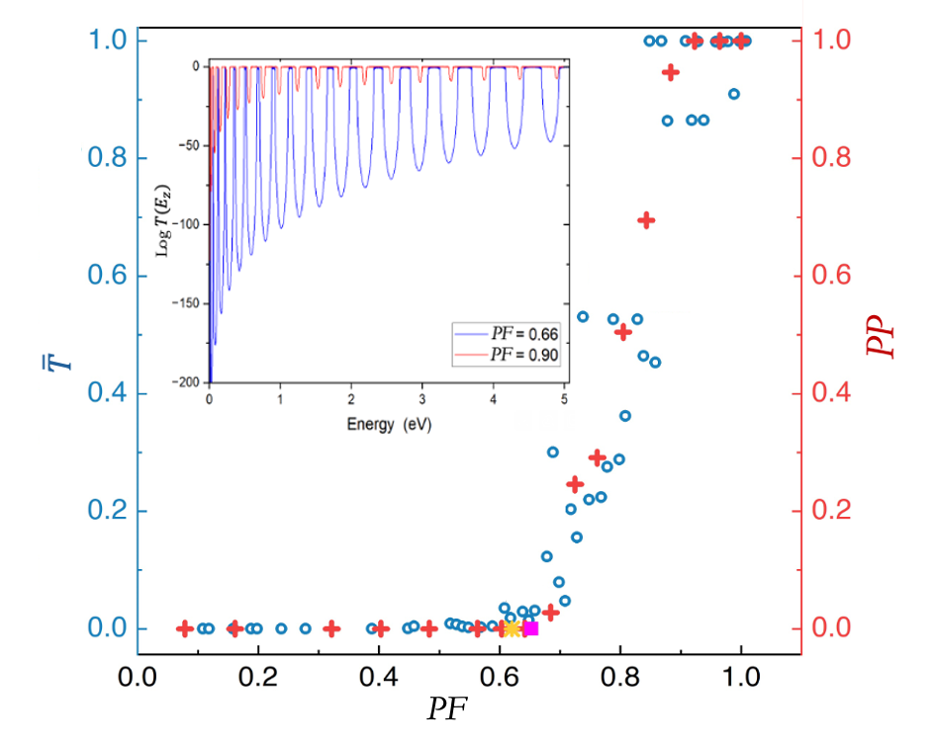}
    \caption{Variation of the transmission probability, $\overline{T}$ (normalized to the maximum $\overline{T}$ of 0.74) through a QD constituted array as a function of the \textit{PF} is shown  \textcolor{RoyalBlue}{{(blue \ding{109} symbols)}} alongside the results of measurements of the percolation probability, \textit{PP} \textcolor{Red}{(red + symbols)} for light injected into waveguides – further described in the main text. The obtained $PF_c$  of $\sim$ 0.64 \textit{i.e.} the \textit{PF} when $\overline{T}$ increases substantially from zero is indicated.  The results of an EIS experiment which suggest a $PF_c$ of $\sim$ 0.62 are also shown in \textcolor{Dandelion}{yellow}, while the analytical solution for $PF_c$ (=0.65) of 2D hexagonal lattices is shown in \textcolor{Plum}{purple}. \textit{Inset:} Log $T(E_z)$ for selected \textit{PFs}. As the \textit{PF} increases, more of the energy levels in the QDs exhibit a unity $T(E_z)$, as indicated by the wider plateaus.}
    \label{3}
\end{figure}

For determining the validity of the obtained \textit{PF\textsubscript{c}}, a direct comparison through experiments would be difficult\textsuperscript{26} – as the precise arrangement\textsuperscript{27} of QDs and vacancies within any given QD array, or over several sample QD arrays, would not be practically easy to determine. However, an instance where the modeled behavior may be applicable was noted through comparison with impedance spectroscopy experiments\textsuperscript{28,29 }performed on QD monolayers assembled through the Langmuir-Blodgett (LB) technique \textsuperscript{30}, in which the \textit{PF} was regulated by mechanical compression of the monolayer. There, it was found that with QDs (of radius \textit{R}) and separated by a distance \textit{s) } in a hexagonal lattice with  $\frac{2R+s}{2R}\leq1.2$ that the electrical current through the QD constituted films exhibited a transition from capacitive to inductive behavior. Here, the \textit{PF }may be derived to be  $\frac{4\pi R^2}{8\sqrt{3}R^2+8\sqrt{3}Rs+2\sqrt{3}s^2}$, and the factor of 1.2 may be related \textit{to }a $PF\geq0.62$ suggesting a close connection to the threshold value presented here.

Another plausibility argument, related to the validity of the proposed transport mechanism through a disordered array, could be made through comparison with an experiment related to the propagation of light in two-dimensional hexagonal lattices constituted from a (40 x 40) lattice point array – with an individual lattice point corresponding to a photonic waveguide\textsuperscript{13}. There, the waveguides were randomly placed,\textsuperscript{ 31 }leading to a variable \textit{PF}, following closely the precepts in our own work. In any given trial, light was injected at a point at the front side of a slab into a central waveguide, subsequent to which the output light intensity was monitored as a function of spatial position on the back side of the slab. As the \textit{PF} of the lattice was increased, the output light intensity measured at non-collinear, but parallel, waveguides was found to increase. The calculated\textsuperscript{13} percolation probability, $PP(=\frac{\sum_{i=1}^{N_p}x_i}{N_p})$ with \textit{i }as the index of the trials, and \textit{N\textsubscript{p} }being the number of trials ($\sim$40) performed at each \textit{PF} is indicated on \textit{the right axis }in Fig. 3. The \textit{x\textsubscript{i} }was set to 1 (/0) if at least 10\% of the input light intensity was (/was not) measured at the output ports in any given trial\textit{. }An\textit{ excellent }agreement in the corresponding \textit{PF\textsubscript{c }}as well as the increase above the \textit{PF\textsubscript{c }}was observed\textit{, }as illustrated in Fig. 3\textbf{. }Even though the length scales between adjacent sites in the QD arrays ($\sim$5 nm) and in the waveguide arrays ($\sim$15 $\mu$m) are distinct, the commonality in terms of a threshold-like behavior in the transport of electrical charges as well as light in a disordered array of confining moieties appears well modeled based on the rationale considered in our work. 

In summary, the modeling of the transmission of charges in dispersed QD assemblies as a function of disorder has been interpreted in terms of traversal through representative potential landscapes and connected to a threshold value of the packing fraction. The consequent aspect of the transition from a non-conducting \textit{to }a conducting state was understood in terms of successful transport paths which span the array. While we have indicated that such a transition in the packing fraction dependent transmission probability ($\overline{T}$) occurs at a critical \textit{PF }of $\sim$ 0.64, the conversion to a measured current still poses a challenge and may involve aspects related to determining the \textit{actual }number of successful paths. The indicated approach is capable of yielding insights into the modality of carrier passage for a broad range of situations \textit{i.e. }involving both charges in QD arrays as well as light in waveguides.

\section*{Citations and References}

 \textsuperscript{1} D. Spittel, J. Poppe, C. Meerbach, C. Ziegler, S.G. Hickey, and A. Eychmüller, Absolute Energy Level Positions in CdSe Nanostructures from Potential-Modulated Absorption Spectroscopy (EMAS), ACS Nano \textbf{11}(12), 12174–12184 (2017).

 \textsuperscript{2} N. Yazdani, S. Andermatt, M. Yarema, V. Farto, M.H. Bani-Hashemian, S. Volk, W.M.M. Lin, O. Yarema, M. Luisier, and V. Wood, Charge transport in semiconductors assembled from nanocrystal quantum dots, Nat Commun \textbf{11}(1), 2852 (2020).

 \textsuperscript{3} J.M. Pietryga, Y.-S. Park, J. Lim, A.F. Fidler, W.K. Bae, S. Brovelli, and V.I. Klimov, Spectroscopic and Device Aspects of Nanocrystal Quantum Dots, Chem Rev \textbf{116}(18), (2016).

 \textsuperscript{4} F. Hetsch, N. Zhao, S. V. Kershaw, and A.L. Rogach, Quantum dot field effect transistors, Materials Today \textbf{16}(9), 312–325 (2013).

 \textsuperscript{5} D. Yu, C. Wang, and P. Guyot-Sionnest, n-Type Conducting CdSe Nanocrystal Solids, Science (1979) \textbf{300}(5623), 1277–1280 (2003).

 \textsuperscript{6} D. Yu, C. Wang, B.L. Wehrenberg, and P. Guyot-Sionnest, Variable Range Hopping Conduction in Semiconductor Nanocrystal Solids, Phys Rev Lett \textbf{92}(21), 216802 (2004).

 \textsuperscript{7} P. Guyot-Sionnest, Electrical Transport in Colloidal Quantum Dot Films, J Phys Chem Lett \textbf{3}(9), 1169–1175 (2012).

 \textsuperscript{8} H.E. Romero, and M. Drndic, Coulomb Blockade and Hopping Conduction in PbSe Quantum Dots, Phys Rev Lett \textbf{95}(15), 156801 (2005).

 \textsuperscript{9} H. Liu, A. Pourret, and P. Guyot-Sionnest, Mott and Efros-Shklovskii Variable Range Hopping in CdSe Quantum Dots Films, ACS Nano \textbf{4}(9), 5211–5216 (2010).

 \textsuperscript{10} D. V. Talapin, J.-S. Lee, M. V. Kovalenko, and E. V. Shevchenko, Prospects of Colloidal Nanocrystals for Electronic and Optoelectronic Applications, Chem Rev \textbf{110}(1), 389–458 (2010).

 \textsuperscript{11} N.S. Makarov, J. Lim, Q. Lin, J.W. Lewellen, N.A. Moody, I. Robel, and J.M. Pietryga, Quantum Dot Thin-Films as Rugged, High-Performance Photocathodes, Nano Lett \textbf{17}(4), 2319–2327 (2017).

 \textsuperscript{12} B.J. Last, and D.J. Thouless, Percolation Theory and Electrical Conductivity, Phys Rev Lett \textbf{27}(25), 1719–1721 (1971).

 \textsuperscript{13} Z. Feng, B.-H. Wu, H. Tang, L.-F. Qiao, X.-W. Wang, X.-Y. Xu, Z.-Q. Jiao, J. Gao, and X.-M. Jin, Direct observation of quantum percolation dynamics, Nanophotonics \textbf{12}(3), 559–567 (2023).

 \textsuperscript{14} J.S. Walker, and J. Gathright, Exploring one‐dimensional quantum mechanics with transfer matrices, Am J Phys \textbf{62}(5), (1994).

 \textsuperscript{15} P. Markos, and C. Soukoulis, \textit{Wave Propagation }(Princeton University Press, 2008).

 \textsuperscript{16} K.L. Jensen, \textit{Introduction to the Physics of Electron Emission}, 1st ed. (Wiley, New York, NY, 2017).

 \textsuperscript{17} M. Mardaani, A.A. Shokri, and K. Esfarjani, Analytical results on coherent conductance in a general periodic quantum dot: Transfer matrix method, Physica E Low Dimens Syst Nanostruct \textbf{28}(2), 150–161 (2005).

 \textsuperscript{18} H. Lepage, A. Kaminski-Cachopo, A. Poncet, and G. le Carval, Simulation of Electronic Transport in Silicon Nanocrystal Solids, The Journal of Physical Chemistry C \textbf{116}(20), 10873–10880 (2012).

 \textsuperscript{19} X. Chu, H. Heidari, A. Abelson, D. Unruh, C. Hansen, C. Qian, G. Zimanyi, M. Law, and A.J. Moulé, Structural characterization of a polycrystalline epitaxially-fused colloidal quantum dot superlattice by electron tomography, J Mater Chem A Mater \textbf{8}(35), 18254–18265 (2020).

 \textsuperscript{20} Y. Xing, N. Yazdani, W.M.M. Lin, M. Yarema, R. Zahn, and V. Wood, Effect of Positional Disorders on Charge Transport in Nanocrystal Quantum Dot Thin Films, ACS Appl Electron Mater \textbf{4}(2), 631–642 (2022).

 \textsuperscript{21} L. Qu, M. Vörös, and G.T. Zimanyi, Metal-Insulator Transition in Nanoparticle Solids: Insights from Kinetic Monte Carlo Simulations, Sci Rep \textbf{7}(1), 7071 (2017).

 \textsuperscript{22} K. Hongo, R. Maezono, and K. Miura, Random number generators tested on quantum Monte Carlo simulations, J Comput Chem \textbf{31}(11), 2186–2194 (2010).

 \textsuperscript{23} Y. Stoyan, and G. Yaskov, Packing congruent hyperspheres into a hypersphere, Journal of Global Optimization \textbf{52}(4), 855–868 (2012).

 \textsuperscript{24} R.W. Rendell, M.G. Ancona, W. Kruppa, E.E. Foos, A.W. Snow, D. Park, and J.B. Boos, Electron transport in nanocluster films with random voids, IEEE Trans Nanotechnol \textbf{2}(2), 75–81 (2003).

 \textsuperscript{25} F. Yonezawa, S. Sakamoto, and M. Hori, Percolation in two-dimensional lattices. I. A technique for the estimation of thresholds, Phys Rev B \textbf{40}(1), 636–649 (1989).

 \textsuperscript{26} H. Lan, and Y. Ding, Ordering, positioning and uniformity of quantum dot arrays, Nano Today \textbf{7}(2), 94–123 (2012).

 \textsuperscript{27} W.Q. Ma, Y.W. Sun, X.J. Yang, D.S. Jiang, and L.H. Chen, Distinct two dimensional lateral ordering of self-assembled quantum dots, Physica E Low Dimens Syst Nanostruct \textbf{40}(6), 1952–1954 (2008).

 \textsuperscript{28} G. Markovich, C.P. Collier, and J.R. Heath, Reversible Metal-Insulator Transition in Ordered Metal Nanocrystal Monolayers Observed by Impedance Spectroscopy, Phys Rev Lett \textbf{80}(17), 3807–3810 (1998).

 \textsuperscript{29} G. Markovich, C.P. Collier, S.E. Henrichs, F. Remacle, R.D. Levine, and J.R. Heath, Architectonic Quantum Dot Solids, Acc Chem Res \textbf{32}(5), 415–423 (1999).

 \textsuperscript{30} O.N. Oliveira, L. Caseli, and K. Ariga, The Past and the Future of Langmuir and Langmuir–Blodgett Films, Chem Rev \textbf{122}(6), 6459–6513 (2022).

 \textsuperscript{31} M. Herrero-Collantes, and J.C. Garcia-Escartin, Quantum random number generators, Rev Mod Phys \textbf{89}(1), 015004 (2017).

\section*{Acknowledgements}

The authors are grateful for financial support from the UCSD- LANL (Los Alamos National Laboratory) initiative courtesy Dr. C. Farrar. and Dr. H. R. Trellue. Discussions with Dr. H. Yamada are much appreciated.

\bibliography{apssamp}

\end{document}